\documentclass{PoS}
\usepackage{amsmath} 
\newcommand{\AUT} {$A^{\sin(\phi-\phi_S)}_{\text{UT}}$}
\newcommand{\AUTphiS} {$A^{\sin\phi_S}_{\text{UT}}$}

\newcommand{\re}{\mathrm{Re}\,}
\newcommand{\im}{\mathrm{Im}\,}

\DeclareSymbolFont{letters}     {OML}{cmm}{m}{it}
\DeclareSymbolFont{symbols}     {OMS}{cmsy}{m}{n}
\DeclareSymbolFont{largesymbols}{OMX}{cmex}{m}{n}

\title{Transverse target spin asymmetries in exclusive $\rho^0$ muoproduction}

\ShortTitle{Transverse target spin asymmetries in exclusive $\rho^0$ muoproduction}

\author{\speaker{Katharina SCHMIDT}\\
        Freiburg University\\
        on behalf of the COMPASS Collaboration\\
        E-mail: \email{Katharina.Schmidt@cern.ch}}

\abstract{Generalized Parton Distributions (GPDs) provide a dynamical picture of the nucleon. The exclusive production of $\rho^0$ mesons on a transversely polarised target is sensitive to the nucleon helicity-flip GPDs $E$ which are related to the total angular momentum of quarks and gluons.\\ 
In 2007 and 2010 the COMPASS experiment at CERN collected data by scattering a 160 GeV/c muon beam off a transversely polarised NH$_3$ target. The final state particles were detected with the two-stage spectrometer with high resolution tracking. In this talk new results for the azimuthal asymmetries $A_{\text{UT}}$ and $A_{\text{LT}}$ are presented.}

\FullConference{XXI International Workshop on Deep-Inelastic Scattering and Related Subjects\\
		 22-26 April, 2013\\
		 Marseilles, France}

\begin{document}

\section{Introduction}
Generalized Parton Distributions (GPDs) provide an extensive description of the internal quark-gluon structure of the nucleon. They contain the well know form factors and the parton distributions functions as sum rules and limiting cases. Experimentally, GPDs can be probed via hard exclusive production of photons (Deep Virtual Compton Scattering, DVCS) or mesons (Hard Exclusive Meson Production, HEMP). Recent reviews can be found in Ref.~\cite{Guidal:2013rya,Kroll:2012sm}.
For exclusive production of mesons by longitudinally polarised photons the factorisation theorem is valid and it allows for a separation of the amplitude into a hard part, described by perturbative QCD and a soft part (Ref.~\cite{Collins:1996fb}). This non-perturbative part enfolds the structure of the nucleon described by GPDs and the structure of the emitted meson depicted by the distribution amplitude.

The measurement of hard exclusive $\rho^0$ mesons is performed at the COMPASS experiment, a fixed-target experiment situated at the SPS M2 beam line at CERN (Ref.~\cite{Abbon:2007pq}).  
A $\mu ^+$ beam with an intensity of $\sim$ 2 $\cdot 10^8$ muons per SPS cycle, a momentum of $\sim$ 160 GeV/$c$ and a longitudinally polarisation of about $P_{\ell} \sim 80$\% is scattered off a transversely polarised NH$_3$ target. The target has three target cells, of 30~cm, 60~cm and 30~cm length each, arranged along the beam axis, were the nucleon polarisation in neighbouring cells are opposite. The achieved target polarisation $P_T$ is routinely larger than $80$\%. The fraction of polarisable material in the target, weighted by the cross section, is quantified by the dilution factor $f$ $\sim 25$\%. The setup allows the measurement of both spin polarisation states at the same time to reduce systematic effects introduced by beam flux variations. Systematic effects due to acceptance variations are further suppressed by rotating the spin direction periodically about once a week.   
\newline
The cross section of hard exclusive $\rho^0$ meson production off transversely polarised protons in the COMPASS kinematic region is given by Ref.~\cite{Diehl:2005pc}:
\begin{align}
  \label{Xsection}
\frac{d\sigma}{dx_B\, dQ^2\, dt\, d\phi\, d\phi_S}
 &= \bigg[ \frac{\alpha_{\rm em}}{8\pi^3}\, 
\frac{y^2}{1-\varepsilon}\,
       \frac{1-x_{Bj}}{x_{Bj}}\, \frac{1}{Q^2} \bigg] \Bigg\{
\frac{1}{2} \Big( \sigma_{++}^{++} + \sigma_{++}^{--} \Big)
+ \varepsilon \sigma_{00}^{++} 
- \varepsilon \cos(2\phi)\, \re \sigma_{+-}^{++} \nonumber \\
& \hspace{1em} {}
- \sqrt{\varepsilon (1+\varepsilon)}\,
  \cos\phi\, \re (\sigma_{+0}^{++} + \sigma_{+0}^{--})
- P_\ell\, \sqrt{\varepsilon (1-\varepsilon)}\, 
           \sin\phi\, \im (\sigma_{+0}^{++} + \sigma_{+0}^{--})
\phantom{\Bigg[ \Bigg] }
\nonumber \\
&- S_T\, \bigg[
  \sin(\phi-\phi_S)\,
  \im (\sigma_{++}^{+-} + \varepsilon \sigma_{00}^{+-})
+ \frac{\varepsilon}{2} \sin(\phi+\phi_S)\, \im \sigma_{+-}^{+-}
\nonumber \\
& \hspace{1em} {}
+ \frac{\varepsilon}{2} \sin(3\phi-\phi_S)\, \im \sigma_{+-}^{-+}
+ \sqrt{\varepsilon (1+\varepsilon)}\, 
  \sin\phi_S\, \im \sigma_{+0}^{+-}
\nonumber \\
& \hspace{1em} {}
+ \sqrt{\varepsilon (1+\varepsilon)}\, 
  \sin(2\phi-\phi_S)\,  \im \sigma_{+0}^{-+}
\bigg]
\nonumber \\
&+ S_T P_\ell\, \bigg[
  \sqrt{1-\varepsilon^2}\, \cos(\phi-\phi_S)\, \re \sigma_{++}^{+-}
- \sqrt{\varepsilon (1-\varepsilon)}\, 
  \cos\phi_S\, \re \sigma_{+0}^{+-}
\nonumber \\
& \hspace{1em} {}
- \sqrt{\varepsilon (1-\varepsilon)}\, 
  \cos(2\phi-\phi_S)\,  \re \sigma_{+0}^{-+} 
\bigg] \Bigg\} ,
\end{align}
where only the terms relevant for this analysis are shown explicitly. 
The azimuthal angle between the lepton scattering plane and the plane containing the virtual photon and the produced meson is denoted by $\phi$, and $\phi _S$ is the azimuthal angle of the transverse component of the target spin vector $S_T$ around the virtual photon direction relative to the lepton scattering plane. The virtual photon polarisation parameter is given by $\epsilon$. The symbols $\sigma_{\mu \sigma}^{\nu\lambda}$ stand for polarised photoabsorption cross sections or interference terms. Experimentally accessible are eight cross-section asymmetries:
\begin{alignat}{5}
  \label{eq:asym_def}
  &A_{\text{UT}}^{\sin(\phi-\phi_s) } &=&   - \frac{\rm{Im}(\sigma_{++}^{+-} + \varepsilon \; \sigma_{00}^{+-})}
  {\sigma_0},
\qquad \qquad
 &A&_{\text{UT}}^{\sin(\phi_s) } &=&   - \frac{\im \sigma_{+0}^{+-}}{\sigma_0}, \nonumber \\
 &A_{\text{UT}}^{\sin(\phi+\phi_s) }  &=& - \frac{\im \sigma_{+-}^{+-}}{\sigma_0}, 
\qquad \qquad
 &A&_{\text{LT}}^{\cos(\phi-\phi_S)}  &=&  \frac{\re \sigma_{++}^{+-}}{\sigma_0},
\nonumber \\
&A_{\text{UT}}^{\sin(2\phi-\phi_s) } &=&   - \frac{\im \sigma_{+0}^{-+}}{\sigma_0},
\qquad \qquad
&A&_{\text{LT}}^{\cos(\phi_s) } &=& \frac{\re \sigma_{+0}^{+-}}{\sigma_0},
\nonumber \\ 
&A_{\text{UT}}^{\sin(3\phi-\phi_s) } &=& - \frac{\im \sigma_{+-}^{-+}}{\sigma_0},
\qquad
&A&_{\text{LT}}^{\cos(2\phi-\phi_s) } &=& \frac{ \re \sigma_{+0}^{-+}}{\sigma_0}.
\nonumber \\
\end{alignat}
Here, $\sigma_0$ is the unpolarised cross section.

\section{Event selection and background estimation}
The analysis is based on all data measured so far by the COMPASS Collaboration utilizing a transversely polarised NH$_3$ target.
Here a short summary of the applied cuts will be given. Details on the event selection can be found in the recent COMPASS publication~\cite{Adolph:2012ht}.
The considered events are characterized by an incoming and a scattered muon and two oppositely charged hadrons associated to a vertex in the polarised target. In order to select a sample in the deep inelastic scattering regime and to keep radiative corrections negligible, the following cuts are used: \\
$Q^2 > $ 1 (GeV/$c$)$^2$, 0.003 $<x_{Bj}<$ 0.35, $W>$ 5 GeV/$c^2$ and  the fractional energy of the virtual photon 0.1$<y<$ 0.9. The $\rho ^0$ meson is selected in the two-hadron invariant mass range\\ 0.5 $< M_{\pi ^+\pi ^-} <$ 1.1 GeV/$c^2$, where for each hadron the pion mass hypothesis is assigned.
The measurements are performed without detection of the recoiling proton in the final state. The exclusivity selection of the events is based on cuts on the missing energy $ |E_{\text{miss}} | <$ 2.5 GeV. Here, $E_{\text{miss}} = \frac{(p+q-v)^2 - p^2}{2 M_P} = \frac{M_X^2 - M_P^2}{2  M_P}$, where $M_X$ the mass of the undetected recoiling system, which is calculated with the four-momenta of proton $p$, photon $q$, and meson $v$. The proton is assumed to be at rest. Non-exclusive background can be suppressed further by the following cuts: The squared transverse momentum of the vector meson with respect to the virtual photon direction\\
 $p_T^2 <$ 0.5 (GeV/$c$)$^2$, the energy of $\rho ^0$ in the laboratory system $E_{\rho ^0} >$ 15 GeV and $Q^2<$ 10 (GeV/$c$)$^2$. A cut on $p_T^2>0.05$ (GeV/$c$)$^2$ is used to reduce coherently produced events. After all cuts applied, the final data set of incoherently produced exclusive $\rho ^0$ events consist of about 797000 events.  The average values of the kinematic variables are $\langle Q^2 \rangle = 2.15 $ (GeV/$c$)$^2$, $\langle x_{Bj} \rangle = 0.039$, $\langle y \rangle = 0.24$, $\langle W \rangle = 8.13$ GeV/$c^2$, and $\langle p_T^2 \rangle = 0.18$ (GeV/$c$)$^2$. 

In order to correct for the remaining semi-inclusive background the $E_{\text{miss}}$ shape of the background is parametrized for each individual target cell in every kinematic bin of $Q^2$, $x_{Bj}$, and $p_{T}^2$ using a Monte-Carlo (MC) LEPTO sample. Prior, the $h^+h^-$ MC event sample is weighted in every $E_{\text{miss}}$ bin  by the ratio of numbers of like-sign events from data and MC.
In every bin $k$ required for the asymmetry extraction, i.e the kinematic bins $Q^2$, $x_{Bj}$, and $p_{T}^2$, individually for each target cell and spin orientation a signal plus background fit is performed, with a Gaussian function for the signal and the background shape fixed by MC as described above. An example can be found in Ref.~\cite{Adolph:2012ht}. 
In order to arrive at background corrected distributions, $N^{\text{sig}}_k(\phi,\phi_S)$, we consider the measured distributions in the signal region, $N^{\text{sig,raw}}_k(\phi,\phi_S)$, and in the background region\\
 7 $< E_{\text{miss}} <$ 20 GeV, $N^{\text{back}}_k(\phi,\phi_S)$. 
The distributions $N^{\text{back}}_k(\phi,\phi_S)$
are rescaled with the estimated numbers of background events in the signal region and afterwards subtracted from the $N^{\text{sig,raw}}_k(\phi,\phi_S)$ distributions.
Finally, the asymmetries are extracted using these corrected $N^{\text{sig}}_k(\phi,\phi_S)$ distributions.  The total amount of semi-inclusive background in the signal range is 22\%. After the described subtraction of semi-inclusive  background the final sample still contains diffractive events where the recoiled nucleon is in an excited  N$^{*}$ or $\Delta$ state (14\%), coherently produced $\rho ^0$ mesons ($\sim$ 5\%), and non-resonant $\pi ^+ \pi ^-$ pairs ($<$ 2\%) (Ref.~\cite{Adolph:2012ht}). We do not apply further corrections for these contributions.

\section{Results}
\begin{figure}
\begin{center}
\begin{minipage}[c]{0.35\textwidth}
\includegraphics[trim=0mm 0mm 0mm 0mm, clip,scale=0.39]{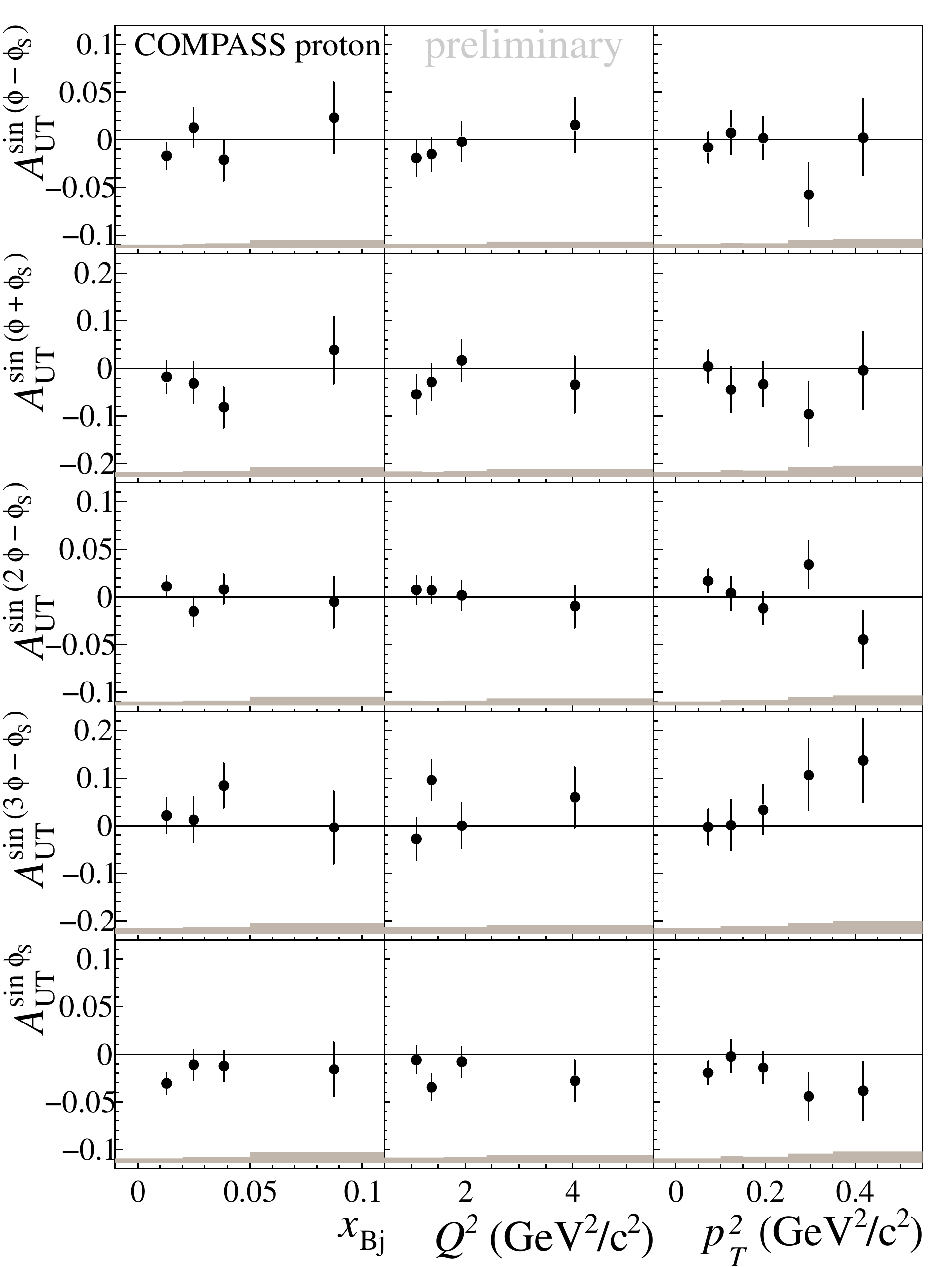}
\end{minipage}
\hfill
\begin{minipage}[c]{0.48\textwidth}
\includegraphics[trim=0mm 0mm 0mm 0mm, clip,scale=0.39]{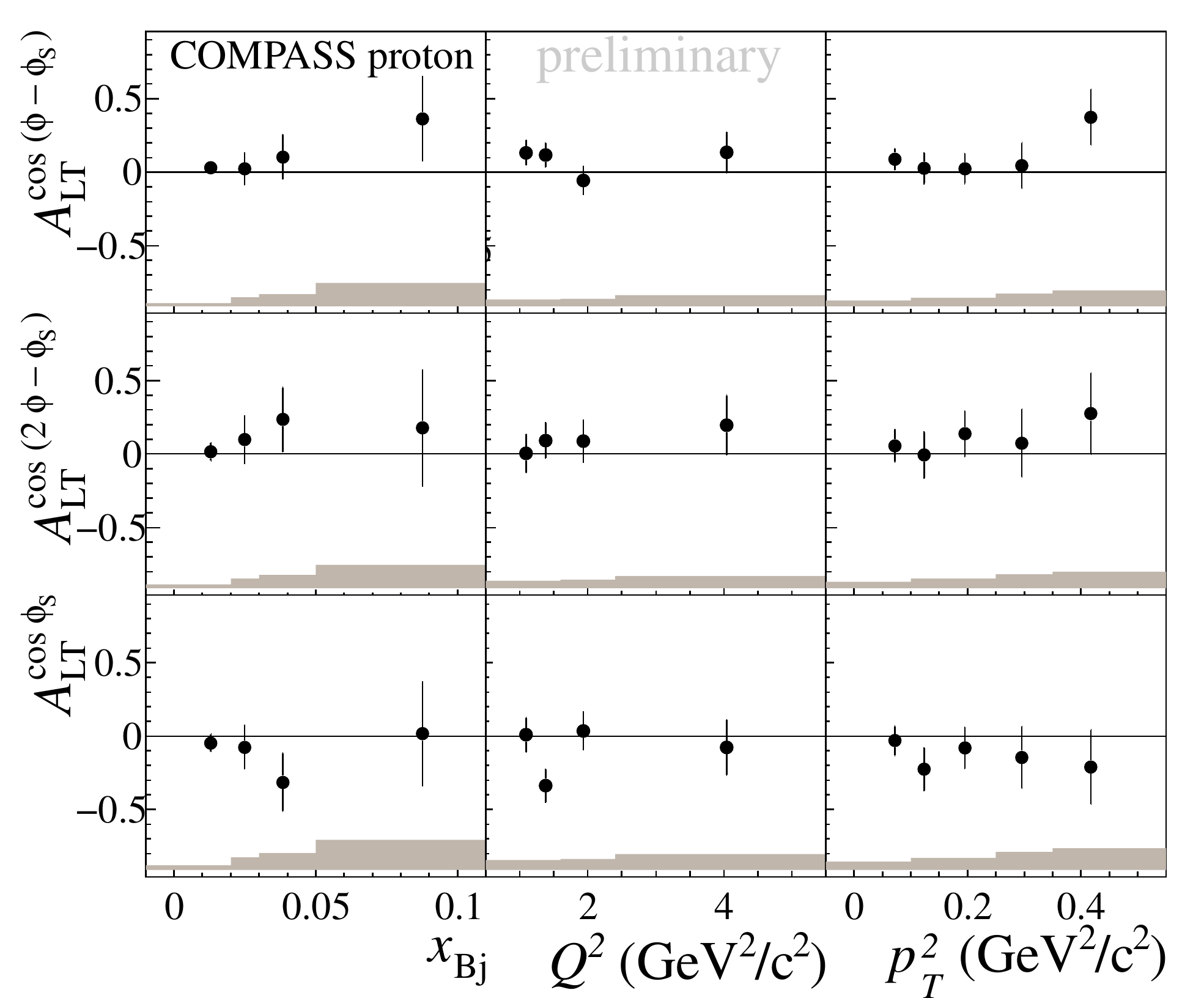}
\end{minipage}
\caption[]{Single-spin azimuthal asymmetries (left side) and double-spin azimuthal asymmetries (right side). The error bars (bands) represent the statistical (systematic) uncertainties.}
\label{fig:asym}
\end{center}\end{figure}

\begin{figure}
\begin{center}
\includegraphics[trim=0mm 0mm 0mm 0mm, clip,scale=0.4]{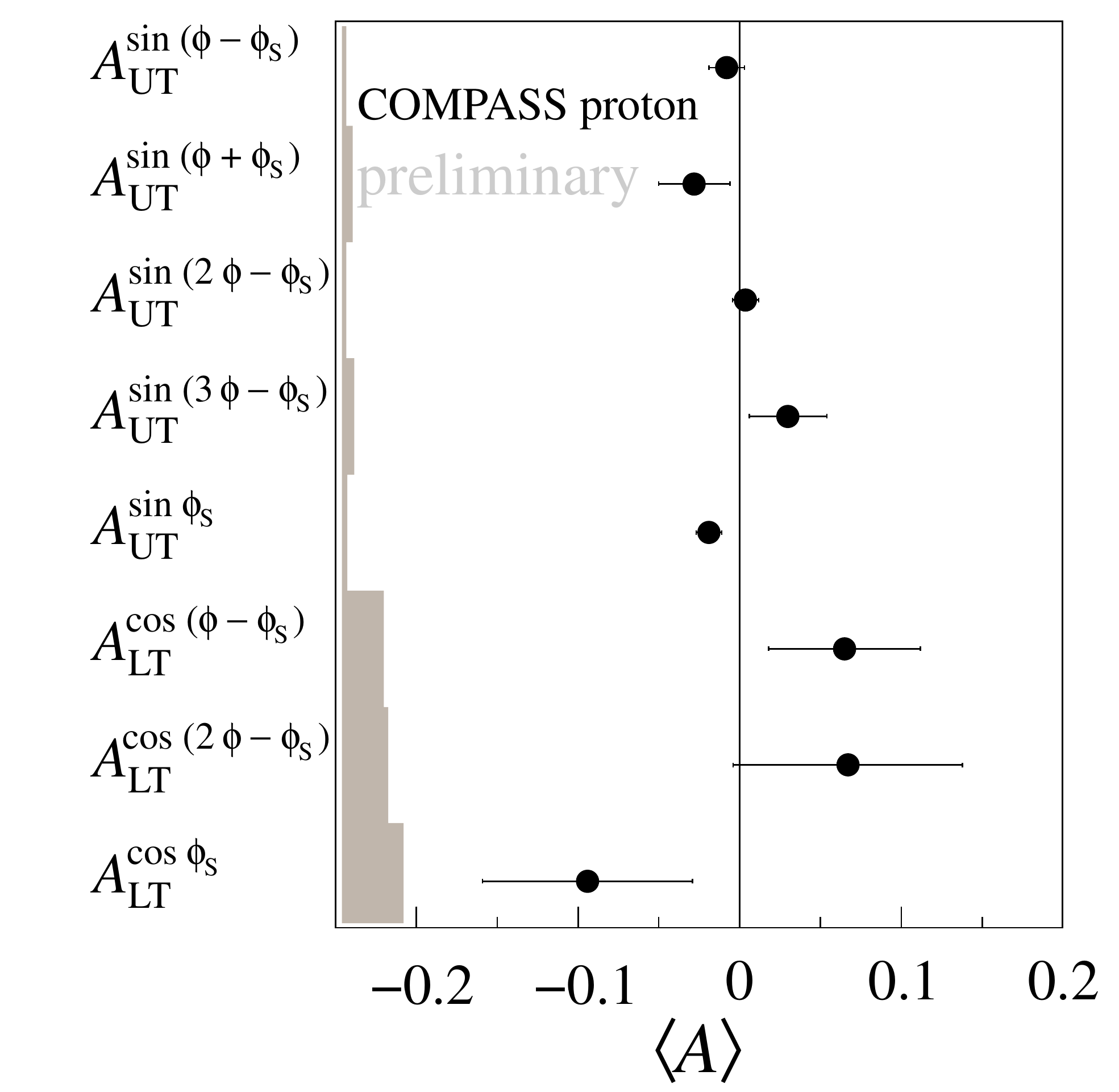}
\caption[]{Mean  value $\langle A \rangle$  for every modulation. The error bars (left bands) represent the statistical (systematic) uncertainties.}
\label{fig:asym_mean}
\end{center}
\end{figure}

The asymmetries are evaluated by combining periods with different target polarisations. Events of the two outer target cells are summed up. The number of exclusive $\rho ^0$ mesons as a function of $\phi$  and $\phi _S$ (with $j$ = number of bins in $\phi$, $\phi _S$) for every target cell ($n$) are given by\\ $N^{\pm} _{j,n} (\phi, \phi_S) = a^{\pm}_{j,n} \left( 1 \pm  A(\phi,\phi_S) \right)$, where $a^{\pm}_{j,n}$ is the  product of the spin-averaged cross section, the muon flux, the number of target nucleons, the acceptance, and the efficiency of the spectrometer. The angular dependence reads:
\begin{eqnarray}
A(\phi, \phi_S) &=& A_{\text{UT,raw}}^{\sin(\phi - \phi _S)} \sin(\phi - \phi _S)+ A_{\text{UT,raw}}^{\sin(\phi+\phi _S)} \sin(\phi+\phi _S) \nonumber \\ 
&& + { } A_{\text{UT,raw}}^{\sin(3\phi -\phi _S)} \sin(3\phi - \phi _S) 
+ A_{\text{UT,raw}}^{\sin(2\phi - \phi _S)} \sin(2\phi - \phi _ S) \nonumber \\
&& + { } A_{\text{UT,raw}}^{\sin(\phi _S)} \sin(\phi _S) + A_{\text{LT,raw}}^{\cos(\phi - \phi _S)} \cos(\phi - \phi _S) \nonumber \\
&& + { } A_{\text{LT,raw}}^{\cos (\phi _S)} \cos(\phi _S) + A_{\text{LT,raw}}^{\cos(2\phi -\phi _S)} \cos(2\phi - \phi _S).
\end{eqnarray}
After the subtraction of semi-inclusive background, all ``raw'' asymmetries $A^m_{\text{UT, raw}}$ and $A^m_{\text{LT,raw}}$ for the different azimuthal modulations $m$ are extracted simultaneously from the final sample  using a two-dimensional binned maximum likelihood fit. They are related to the transverse target asymmetries in Eq.~\ref{eq:asym_def} as:
\begin{eqnarray}
\label{eq:raw_asym}
 A^m_{\text{UT}} &=&  \frac{A^m_{\text{UT,raw}}}{\langle f \cdot |P_{T}| \cdot D^m (\epsilon)\rangle}, \nonumber \\
A^m_{\text{LT}} &=&  \frac{A^m_{\text{LT,raw}}}{\langle f \cdot |P_{T}| \cdot P_{\ell} \cdot D^m (\epsilon) \rangle} .
\end{eqnarray}
The factors $D^m(\epsilon)$ for the different modulations $m$ are given in Ref.~\cite{Diehl:2005pc}:
\begin{alignat} {2}
\label{eq:dnn}
&D^{\sin(\phi - \phi _S)}  &=& 1, \nonumber \\
&D^{\sin(\phi + \phi _S)}  &=&{} D^{\sin(3\phi - \phi _S)}  = \frac{\varepsilon}{2} \approx \frac{1-y}{1+(1-y)^2}, \nonumber \\
&D^{\sin(\phi _S)}  &=& {}D^{\sin(2\phi - \phi _S)} = \sqrt{\varepsilon(1+\varepsilon)} \approx \frac{(2-y) \sqrt{2(1-y)}}{1+(1-y)^2}, \nonumber \\
&D^{\cos(\phi -\phi _S)}  &=& \sqrt{1-\varepsilon^2} \approx \frac{y(2-y)}{1+(1-y)^2}, \nonumber \\
&D^{\cos\phi _S} &=& D^{\cos(2\phi - \phi _S)} = \sqrt{\varepsilon(1-\varepsilon)} \approx \frac{y\sqrt{2(1-y)}}{1+(1-y)^2} .
\end{alignat}
The systematic uncertainties are evaluated taking into account the following contributions: a possible bias of the applied estimator, the stability of the asymmetries over data-taking time, and the robustness of the applied background subtraction method.
 Additionally, we consider the systematic uncertainty of the target dilution factor (2\%), the target polarisation (3\%), and the beam polarisation (5\%). 
 The results for the five single-spin asymmetries and three double-spin asymmetries as a function of $x_{Bj}$, $Q^2$, and $p_T^2$ are shown in Fig.~\ref{fig:asym}. Error bars show statistical uncertainties $\sigma ^{stat}$. The systematic uncertainties $\sigma^{sys}$ are represented by grey shaded bands. Additionally the mean asymmetries are shown in Fig.~\ref{fig:asym_mean}. 

Most amplitudes of the extracted asymmetries are small and consistent with zero over the whole kinematic range. The result for \AUT is in good agreement with the stand-alone extraction recently published in Ref.~\cite{Adolph:2012ht}. \\
The asymmetry \AUTphiS is found to be $-0.019 \pm 0.008 (stat) \pm 0.003 (sys)$. This findings can be interpreted in terms of a possible contribution from chiral-odd, transverse GPDs, as previously introduced in Ref.~\cite{Goloskokov:2009ia} to explain the HERMES results for exclusive $\pi ^+$ production on a transversely polarised proton target. 

\newpage
\section{Conclusion}
Preliminary results for five single-spin and three double-spin azimuthal asymmetries for exclusive $\rho^0$ mesons as a function of $Q^2$, $x_{Bj}$ and $p_{T}^{2}$ are presented. The results are extracted using all transversely polarised proton data from the COMPASS experiment. Most amplitudes of the extracted modulations are small and consistent with zero over the whole kinematic range. The asymmetry $A_{\text{UT}}^{\sin(\phi_S)}$ seems to indicate a non-vanishing value within the experimental uncertainties. By comparing our results to future calculations similar to those in Ref.~\cite{Goloskokov:2009ia} the existence of chiral-odd, transverse GPDs could be fortified.

\end{document}